\documentclass[11pt]{article}
\pdfoutput=1
\usepackage{jheppubmod,amsmath,amssymb}
\usepackage{color}
\usepackage{amsmath,amssymb}
\usepackage{comment}
\usepackage{braket}
\usepackage{mathtools}
\usepackage{psfrag}
\usepackage{array}
\usepackage{amssymb}
\usepackage{amsmath}
\usepackage{amsthm} 
\usepackage{graphicx}
\usepackage{subcaption}
\usepackage{epstopdf}
	
\usepackage{color}
\usepackage{epsfig}
\usepackage[punctsep]{collref}
\collectsep[]{;}	

\def\pb[#1,#2]{\{#1, #2\}}
\def\deb[#1,#2]{[#1,#2]_{\text{D.B.}}}

\def\tO{\widetilde{\cal O}}

\def\Or[#1]{{\text{O}}\left({#1}\right)}
\def\dotl[#1,#2]{\left\langle #1,\, #2 \right\rangle}
\def\dotlb[#1,#2]{\left\langle #1,\, #2 \right\rangle}
\def\dotlm[#1,#2]{\left[ #1,\, #2 \right]}
\def\dotp[#1,#2]{(\vect{#1} \cdot\vect{#2})}
\def\aff[#1,#2]{\hat{#1}(#2)}

\def\n4sym{{\cal N}=4 SYM}
\def\>{\rangle}
\def\<{\langle}
\def\weight[#1,#2,#3]{\{(#1),#2,#3\}}
\def\ads[#1]{$\text{AdS}_{#1}$}

\def\tarelr[#1]{\widetilde{a}^{\text{rel}}_{R#1}}
\def\Oright[#1]{{\cal O}_{R#1}}
\def\Oleft[#1]{{\cal O}_{L#1}}
\def\aleft[#1]{a_{L#1}}
\def\arelr[#1]{a^{\text{rel}}_{R#1}}

\newcommand{\be}{\begin{equation}}
\newcommand{\ee}{\end{equation}}
\newcommand{\ba}{\begin{align}}
\newcommand{\ea}{\end{align}}
\newcommand{\bs}{\begin{split}}
\def\sess\end{split}

\newcommand{\vect}[1]{{#1}}
\newcommand{\norm}[1]{|{\boldsymbol{#1}}|}
\def\tO{\widetilde{\cal O}}

\def\rsz{|\Psi_{0}\rangle}

\def\Tr{{\rm Tr}}
\def\tO{\widetilde{\cal O}}

\title{\LARGE \bf
Black Hole State Dependence as a Single Parameter
}

 \author{Rik van Breukelen}
 \emailAdd{rik.van.breukelen@cern.ch}
 \affiliation{Theoretical Physics Department, CERN, CH-1211 Geneva 23, Switzerland 
 \vspace{2mm}}
 \affiliation{Geneva University, 24 quai Ernest-Ansermet, CH-1214 Geneva 4, Switzerland}

\abstract{
It has previously been proposed that the black hole interior of typical state large black holes in AdS can be described using state-dependent operators. We investigate the possibility that the interior can be described by explicit time dependence, which reduces the state-dependence of the interior operators to a single parameter. We also propose to use the natural cone, obtained from Tomita-Takesaki theory, as a candidate construction for the interior operators.}

\begin{document}

\maketitle

\section{Introduction}

The quantum mechanical behavior of black holes is an ongoing topic started by Hawking \cite{Hawking:1974sw}. The Firewall paradoxes of AMPS \cite{Mathur:2009hf,Almheiri:2013hfa,Marolf:2013dba} have formalized the problems surrounding the quantum mechanics of black holes. Moreover, these problems persist for large black holes in AdS. The AdS/CFT correspondence \cite{Maldacena:1997re}, therefore, provides a powerful tool to study this topic.

The naive conclusion of these paradoxes is that there can be no operators in the CFT describing the interior of the black hole, and that, therefore, the horizon of the black hole is not smooth. It is, however, possible to construct interior operators that depend on the state of the black hole \cite{Papadodimas:2012aq,Papadodimas:2013b,Papadodimas:2013}, and thus have a smooth horizon. The geometry described by these operators contains part of the extended AdS-Schwarzschild geometry, including part of the region beyond the interior \cite{deBoer:2018ibj,deBoer:2019kyr}. In this region it is obvious that time moves opposite to the Killing isometry. The interior operators must, therefore, be explicitly time dependent.

In this paper, we will examine whether the state-dependence of the interior operators can be captured in a simple form by time dependence. State dependence can directly be rewritten as time evolution in the case of the thermofield double state and its time-shifted cousins \cite{Papadodimas:2015xma,vanBreukelen:2017dul}. This is possible, more generally, for any ergodic system, because most states will become equal to the other states under time evolution for an ergotic system. State dependence is, therefore, equal to waiting the appropriate amount of time. We explore whether something similar can happen for typical pure state large black holes in AdS, i.e. whether explicit time dependence is enough to avoid the firewall paradoxes.

We also investigate a candidate construction for the interior operators. Tomita-Takesaki theory was used as a motivation for the construction of the state-dependent interior operators. We continue on this path by using the natural cone, described by Tomita-Takesaki theory, which has the elegant property that the interior operators are the same for all states in the natural cone. We investigate whether the natural cone together with explicit time dependence is enough to describe the interior operators of most typical states.

This paper is organized as follows: In section \ref{sec:intro}, we will discuss explicit time dependence in the case of the thermofield double state, the basics of Tomita-Takesaki theory, and the construction of the state-dependent interior operators called the mirror operators. Next, in section \ref{sec:cplus}, we will examine the various firewall paradoxes and see how explicit time dependence can avoid them. Finally, in section \ref{sec:natcone}, the natural cone is proposed as a candidate construction.

\section{Explicit Time Dependence}
\label{sec:intro}
Explicit time dependence is needed to get a consistent description of the black hole interior. In the following section, we will show this for the eternal black hole and typical pure state black holes. We will also discuss the Tomita-Takesaki construction used to describe the interior operators.

\subsection{Eternal Black Hole}
\begin{figure}
\centering
\begin{subfigure}[t]{0.35\textwidth}
    \includegraphics[width=.8\textwidth]{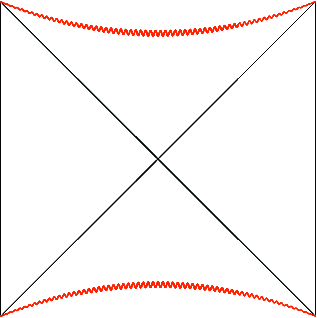}
\caption{\quad \quad \quad \;}
\label{fig:TFD1}
\end{subfigure}
\begin{subfigure}[t]{0.35\textwidth}
    \includegraphics[width=.8\textwidth]{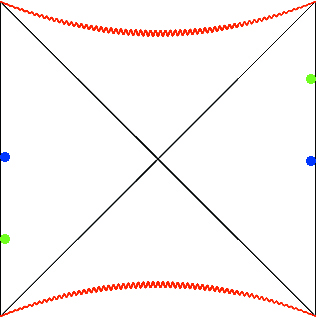}
\caption{\quad \quad \quad}
\label{fig:TFD2}
\end{subfigure}
\caption{a) The Penrose Diagram of the eternal black hole. b) The blue dots correspond to operators which are entangled. An operator at a later time, green dot, is entangled with an earlier operator on the other side.}
\label{fig:TFD}
\end{figure}

The eternal black hole is proposed \cite{Maldacena:2001kr} to be dual to two CFTs in a specific entangled state, the thermofield double state
\begin{equation}
\ket{\Psi_{\text{TFD}}} = \sum_i \frac{e^{-\beta E_i/2}} {\sqrt{Z}} \ket{E_i}_L \ket{E_i}_R,
\end{equation}
where we denote the two CFTs with $(L)$ left and $(R)$ right, and $\beta$ is the inverse temperature of the black hole. The geometry corresponding to this sate is depicted in figure \ref{fig:TFD1}. The geometry has a Killing isometry corresponding to $H_R-H_L$, which flows up on the right part of the geometry and down on the left part of the geometry. This causes the left operator that is entangled with a right operator to move down when we move the right operator up, as depicted in figure \ref{fig:TFD2}. 

To avoid closed timelike curves, we need to impose that time evolution is generated by $H_R+H_L$, i.e. up in both the left and right side of the geometry. This has consequences for left-right correlators,
\begin{equation}
\langle {\cal O}_L(t_1){\cal O}_R(t_2) \rangle = f(t_1+t_2),
\end{equation}
where $f$ is some function of $t_1+t_2$. This is different from the one-sided two-point function, which is a function of $t_1-t_2$. Left and right operators commute, even though the left-right two-point function is non-zero. This is because the two CFTs are causally disconnected. However, by considering a double trace perturbation \cite{Gao:2016bin,Maldacena:2017axo}, of the form $\delta H = g{\cal O}_L{\cal O}_R$, a message can be send from one CFT to the other. This provides evidence for the smoothness of the black hole horizon in this state, as such a probe crosses the horizon.

There is a class of states \cite{Papadodimas:2015xma} that have the same entanglement structure called the time-shifted thermofield double states, 
\begin{equation}
\ket{\Psi_T} = \sum_i \frac{e^{-\beta E_i/2}} {\sqrt{Z}} e^{iE_iT} \ket{E_i}_L \ket{E_i}_R.
\end{equation}
These states can also be made traversable \cite{vanBreukelen:2017dul}, i.e. send a message from one side to other. However, the protocol depends on $T$ and is, therefore, state dependent with a single parameter.

\subsection{Tomita-Takesaki Theory}
\label{sec:TTC}
The construction developed by Tomita and Takesaki state that for a given algebra ${\cal A}$ and reference state $\ket{\psi}$, which obeys
\begin{itemize}
\item $\ket{\psi}$ is cyclic, i.e. ${\cal A}\ket{\psi}$ spans the entire Hilbert space,
\item $\ket{\psi}$ is seperating, i.e. $A\ket{\psi}=0$ only for $A=0$,
\end{itemize}
then the commutant ${\cal A}'$ of the algebra ${\cal A}$ can constructed. This is done by using the Tomita operator $S$ as follows
\begin{align}
 \label{tomita}
\begin{split}
 &\; S A \rsz = A^\dagger \rsz, \\
\Delta &= S^\dagger S\;, \quad \; \; \; S=J\Delta^{1/2}\;,\\
 \widetilde{\cal O}&=J{\cal O}J\;, \quad {\cal A}'=J{\cal A}J\;,
\end{split}
\end{align}
where $J$ is an anti-unitary operator called \textit{modular conjugation}, while $\Delta$ is called the \textit{modular operator}. An in-depth discussion about the properties of Tomita-Takesaki theory can be found in \cite{Bratteli:1979tw}.

It is useful to put this construction in the context of the thermofield double, where we have the following identities
\begin{align}
\begin{split}
{\cal O}_L &= J{\cal O}_R J ,\\
\Delta &= e^{-\beta (H_R-H_L)} .
\end{split}
\end{align}
Here, we notice that the modular operator $\Delta$ is a function $H_R-H_L$, not $H_R+H_L$. This means that we can generate time evolution  on the right exterior with the modular operator by using $\Delta^{is}$, but we need to use explicit time dependence, to compensate for the minus sign, if we also use the modular operator on the left exterior in order to keep a consistent causal structure. This may seem unnecessarily complicated for the thermofield double, however, this is unavoidable in the case of a typical state.

The algebra and its commutant remain the same in the case of the time-shifted thermofield double states. This, however, means that left-right correlation functions change as function of the timeshift. It would, therefore, be necessary to work with precursors that absorb this timeshift to get the same correlation functions as in the situation without timeshift. For example, if we want the same correlators between left and right operators in the thermofield double state and a time-shifted thermofield double state
\begin{equation}
\label{eqn:shift}
\bra{\Psi_{\text{TFD}}} {\cal O}_L(t_1)  {\cal O}_R(t_2) \ket{\Psi_{\text{TFD}}} = \bra{\Psi_T} X_L(t_1)  {\cal O}_R(t_2) \ket{\Psi_T} ,
\end{equation}
we would need to use a precursor $X_L$ on the left side, and identify $X_L(t)={\cal O}_L(t-T)$.

\subsection{Typical Black Hole}
The typical black hole microstate is defined as a superposition of energy eigenstates
\begin{equation}
\ket{\Psi} = \sum c_i \ket{E_i} ,
\end{equation}
where we sum over the energy eigenstates in the window $E_0 \pm \delta E$, and $c_i$ are random complex numbers chosen with the Haar measure. $E_0$ is given by the mass of the black hole and $\delta E$ is an order one number making the window wide enough to account for the entropy of the black hole. Black holes of this type are not formed by normal collapse \cite{Papadodimas:2017qit}.

We want to use the Tomita-Takesaki construction to describe the interior of these black holes. However, typical states are not cyclic and seperating. Nonetheless, they are almost cyclic and almost seperating, i.e. it is difficult to annihilate these states and perturbations around the black hole can be well described by low-point correlation functions. We can, therefore, use the mirror operators constructed in \cite{Papadodimas:2012aq,Papadodimas:2013b,Papadodimas:2013}.

\label{sec:fourier}
We can use a truncated algebra ${\cal A}$ of single-trace operators in frequency space, defined by
\begin{equation}
{\cal A} = {\rm span} \left\{{\cal O}_{\omega_1}, {\cal O}_{\omega_1}{\cal O}_{\omega_2},...., {\cal O}_{\omega_1}....{\cal O}_{\omega_n}\right\} ,
\end{equation}
where $n \ll N$. This linear set is only approximately an algebra as we demand $n\ll N$. Moreover, we will have to work with coarse grained operators to avoid problems caused by the discrete nature of the energy levels. 
\begin{align}
\label{eqn:fourier}
\begin{split}
{\cal O}^{\rm exact}_\omega &\equiv \int_{-\infty}^{+\infty} dt\,\,e^{i \omega t} {\cal O}(t) . \\
{\cal O}_\omega &\equiv {1\over \sqrt{\delta \omega}}\int_{\omega}^{\omega +\delta\omega} {\cal O}^{\rm exact}_{\omega'} d\omega' ,
\end{split}
\end{align}
 where now the set of allowed $\omega$'s is discretized with step $\delta\omega$. In \eqref{eqn:fourier} we have divided by $\sqrt{\delta \omega}$ in order to have an operator whose correlators are stable under small changes of the bin size $\delta \omega$. We also need to impose an upper cutoff in the allowed frequencies $|\omega|\leq \omega_*$. The reason is that the mirror operators are meaningful when the small algebra cannot annihilate the state. In a thermal state we find that $\langle {\cal O}_\omega^\dagger {\cal O}_\omega\rangle \propto e^{-\beta \omega}$. For large $\omega$ this is extremely close to zero, implying that the operator ${\cal O}_\omega$ almost annihilates the state.

The limitations for the algebra are chosen in such a way that typical states are not annihilated by the algebra, and typical states are, therefore, seperating.

The algebra forms, when acting on the reference state, the small Hilbert space
\begin{equation}
{\cal H}_{\ket{\Psi_0}} = {\cal A} \ket{\Psi_0} ,
\end{equation}
in which physics is described with the black hole as a background. It does describe objects falling into the black hole, but not large perturbations to the black hole, such as black hole mergers for example. The typical state is cyclic by construction with respect to the small Hilbert space. Therefore, we can make use of Tomita-Takesaki theory. Moreover, using large $N$ factorization and the KMS condition relevant for equilibrium states, it is possible to show \cite{Papadodimas:2013} that at large $N$ the CFT Hamiltonian acts on the code subspace similar to the (full) modular Hamiltonian 
\be
\label{modularh}
\Delta = \exp[-\beta(H-E_0)] + O(1/N) .
\ee

From \eqref{tomita},\eqref{modularh} follows that at large $N$ the mirror operators are defined by the equations
\begin{align}
\label{defmirror}
\begin{split}
\tO_\omega \rsz &= e^{-{\beta H \over 2}} {\cal O}_\omega^\dagger  e^{{\beta H \over 2}} \rsz ,\\
\tO_\omega {\cal O}_{\omega_1}...{\cal O}_{\omega_n} \rsz &= {\cal O}_{\omega_1}...{\cal O}_{\omega_n} \tO_\omega \rsz ,\\
[H,\tO_\omega]{\cal O}_{\omega_1}...{\cal O}_{\omega_n} \rsz &= \omega \tO_\omega {\cal O}_{\omega_1}...{\cal O}_{\omega_n} \rsz .
\end{split}
\end{align}
The last line generalizes to higher powers of the Hamiltonian $H$, even though $H$ is not in the small algebra.

\label{ssec:mirpos}
Specifying the action of the operators on the small subspace ${\cal H}_{\rsz}$, as in \eqref{defmirror}, is not sufficient to know the time when the operators act. We also need to specify a {\it time ordering} between them and with the normal operators. To describe effective field theory in the bulk, we must define the mirror operators to go against the killing isometry in the bulk. This does not fully fix the time evolution as an overall shift remains. This shift is similar to the relation between the thermofield double and its time-shifted cousins. The one-parameter family, labeled by $T$, of possible choices for how to localize the mirrors in physical time is given by 
\be
\label{mirrortimes}
\tO_T(t) = \int_{-\omega_*}^{\omega_*} d\omega\,\, e^{-i \omega (t-T)}  \tO_\omega ,
\ee
where $t$ labels the physical CFT time at which the operator is localized. Using this Fourier transform, we can setup the mirror operator equations in position space.

\begin{align}
\label{eqn:mp1}
\begin{split}
\widetilde{{\cal O}}_T (t) \ket{\Psi_0} =& \; e^{-\frac{\beta H} {2}} \mathcal{O}^\dagger(T-t) e^{\frac{\beta H} {2}} \ket{\Psi_0} , \\
\tO_T(t) A(t_1,t_2,...) \ket{\Psi_0} =& \; A(t_1,t_2,...) \tO_T(t) \ket{\Psi_0} , \\
[H,\tO_T(t)] A(t_1,t_2,...) \ket{\Psi_0} =& \;  A(t_1,t_2,..) e^{-\frac{\beta H} {2}}[H, \mathcal{O}^\dagger(T-t)] e^{\frac{\beta H} {2}} \ket{\Psi_0} .
\end{split}
\end{align}
These statements are only approximately true and not completely local, as we have binned the Fourier modes and imposed cut-off frequencies. The free parameter $T$ is needed to get a consistent picture for when the mirror operators are defined at different times, i.e. one person uses $\ket{\Psi_0}$ the state, but someone else might set up the experiment some time later and use $e^{iHt} \ket{\Psi_0}$. They must choose $T$ carefully to get consistent results, as the mirror operators they naively obtain -- without $T$ -- are not the same but related as precursors to each other. Normally, it is the complexity that tells you whether you are using the proper operator\footnote{For a bulk picture interpretation one wants to use the simplest operator. Mathematically speaking all precursors are proper operators.} or a precursor. 
We will suppress writing the subscript $T$ for the case $T=0$, such that $\tO_0=\tO$, for notational ease. The results for the case of general $T$ can be obtained as precursors of the $T=0$ case.

\begin{figure}
\centering
\begin{subfigure}[t]{0.35\textwidth}
    \includegraphics[width=.8\textwidth]{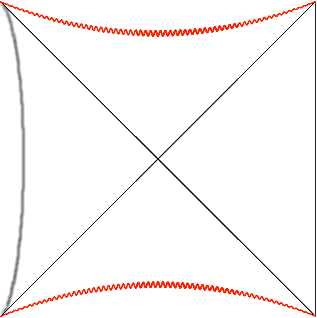}
\end{subfigure}
\begin{subfigure}[t]{0.35\textwidth}
    \includegraphics[width=.8\textwidth]{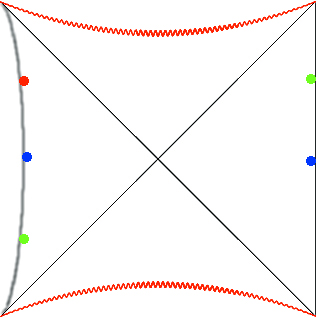}
\end{subfigure}
\caption{a) Penrose diagram of the typical black hole. b) Blue dots denote entangled operators, and green dots denote entangled operators at a later time, similar to the eternal black hole. However, someone using the state at the later time as the reference state for the mirror operator construction would place the location of the operator that is entangled with the right green operator at the red dot, thereby making the construction time-dependent.}
\label{fig:SSBH}
\end{figure}

From the first line of equation \eqref{eqn:mp1}, it is clear that the mirror operators move backwards in time under normal Hamiltonian evolution. This can also be seen from third line of equation \eqref{defmirror}, where the sign is different from what one expects for a normal operator. The mirror operators must, therefore, be explicit time dependent to ensure that they are forward moving in time and to obtain a consistent causal structure.

The geometry of typical state black holes is conjectured \cite{deBoer:2018ibj,deBoer:2019kyr} to include part of the left exterior of the extended Penrose diagram, as depicted in figure \ref{fig:SSBH}. This part of the geometry is described by the mirror operators.

\section{Avoiding the Paradoxes}
\label{sec:cplus}
We have seen that explicit time dependence is a consequence of physical requirements, i.e. to have consistent bulk picture. Explicit time dependence encodes some information about the state in the operators used to construct the interior of a black hole. They are, therefore, state-dependent in this simple form. This leads us to the following claim:
 \vspace{2mm}
\begin{quote}
\centering
\textit{ Explicit time dependence avoids the firewall paradoxes.}
\end{quote}
 \vspace{2mm}
To test the claim, we only need to study a set of states $C$,
\begin{equation}
C=\{\ket{\Psi_1}, \ket{\Psi_2},\dots\},
\end{equation}
where the states $\ket{\Psi_i}, \ket{\Psi_j}$ do not evolve into each other under time evolution, and consider the interior operators as fixed (state-independent) for this set of states. Each state drawn from the Haar measure, for a given energy window, should be part of this set at some time. The interior operators for a given state $\ket{\Psi(0)}$ are thus obtained by finding the time when $\ket{\Psi(t)}$ is part of the set $C$ and use explicit time evolution with the fixed interior operators to obtain the interior operators for $\ket{\Psi(0)}$.

In this section we discuss the various paradoxes \cite{Almheiri:2013hfa,Papadodimas:2013}, and how they are resolved by explicit time dependence.

\subsection{The ``$N_a \neq 0$" paradox}
These arguments \cite{Marolf:2013dba,Bousso:2013wia} support the idea that the horizon is not smooth by comparing the excitations in the modes of the Hawking radiation with the modes of an infalling observer. 

\subsubsection{The Paradox}
The number operator, $N_b = b^\dagger b$, measures the number of particles at frequency $\omega$ seen by the outside
observer. A different number operator,
\begin{equation}
N_a = \frac{1} {1-e^{-\omega \beta}} \left [ \left(b^\dagger - e^{-\omega \beta/2} \widetilde{b}\right)\left(b - e^{-\omega \beta/2} \widetilde{b}^\dagger\right)+\left( \widetilde{b}^\dagger- e^{-\omega \beta/2} b^\dagger\right)\left(\widetilde{b} - e^{-\omega \beta/2}b^\dagger\right)\right] ,
\end{equation}
measures the number of particles seen by the infalling observer. It should be obvious that, to leading order, the following holds
\begin{equation}
[H_{\text{CFT}} , N_b] = 0 ,
\end{equation}
which means that the Hamiltonian and this number operator can be simultaneously diagonalized, at least to leading order.
\begin{align}
\begin{split}
H_{\text{CFT}} \ket{E_i,n_{i,b}} &= E_i \ket{E_i,n_{i,b}} , \\
N_b \ket{E_i,n_{i,b}} &= n_{i,b} \ket{E_i,n_{i,b}} .\\
\end{split}
\end{align}
This leads to a contradiction with the idea that the horizon is smooth when we assume that $N_a$ is a fixed state-independent operator.
\begin{equation}
\label{eqn:numa}
\langle N_a \rangle =  \Tr[\rho_m N_a]/{\cal N} = \bra{E_i,n_{i,b}} N_a \ket{E_i,n_{i,b}} = O(1) ,
\end{equation}
where $\rho_m$ is microcanonical density matrix, and ${\cal N}=\Tr[\rho_m]$ is the normalization. The last equality holds, because $N_a$ is a positive operator and any state with $\langle N_a \rangle = 0$ has a thermal expectation value of $N_b$ and, therefore, the expectation value of $N_a$ must be order one in eigenstates of $N_b$. 

\subsubsection{The Resolution}
\label{ssec:lipr}
\label{sec:lloyds}

It is important that correlators of typical states are close to correlators in the microcanonical ensemble for the previous argument. This is obviously not true in the case of state-dependent operators. In our case the operators are not fully dependent on the state, only on the time. Therefore, it not obvious that the arguments of the paradoxes fail for explicit time dependence.  

For correlators with normal operators averaging over the typical states should give the correlator in the microcanonical ensemble \cite{lloyd},
\begin{equation}
\label{eqn:Lloyd}
\int [d\mu] \bra{\Psi} A \ket{\Psi} = \Tr[\rho_m A]/{\cal N} ,
\end{equation}
Where $A$ is some hermitian operator. We can estimate that each individual typical state is close to microcanonical ensemble by calculating the variance. These calculations are done by averaging over unitaries that rotate the state. This causes the crossterms with random phases to drop out and the diagonal terms remain, which give the trace. 

The measure $[d\mu]$ is used to average over all typical states. It can be decomposed into a part that averages over the size of coefficients $c_i$ of the state, and a part that averages over the phases of the state, $[d\mu]=[d\hat{\mu}] \frac{dt} {t_f}$. The second part, averaging over the phases, is equal to averaging over the time evolution of a state. Therefore, we obtain
\begin{align}
\begin{split}
\Tr[\rho_m A]/{\cal N} &= \int [d\mu] \bra{\Psi} A \ket{\Psi}  ,\\
&= \int [d\hat{\mu}] \int_0^{t_f} \frac{dt} {t_f} \bra{\Psi(t)} A \ket{\Psi(t)}  ,\\
&= \int [d\hat{\mu}] \langle A \rangle_\Psi ,
\end{split}
\end{align}
where $t_f$ is the time of a full orbit, and $\langle A \rangle_\Psi $ is the long time average. 

We can use the same steps to get an estimate for the variance to conclude that most states in the energy window are close to the microcanonical ensemble. However, when $A$ is partially constructed from interior operators, this construction fails, because the interior operators introduce an explicit time dependence and we cannot use the third line. 

The correlator of a time-independent operator is close to the long time average, averaging only over the time of the state. The explicit time dependence of the interior operators means that we cannot use equation \eqref{eqn:numa} for a typical state as we cannot do the change of basis while we have both normal and interior operators in the correlator. Correlators with both normal and interior operators are not close to the micro-canonical ensemble, and the paradox is, therefore, avoided.

It is important to note that the interior Fourier modes are sensitive to the explicit time dependence, even though they act like Fourier modes. Firstly, the interior modes only need to act like Fourier modes in the given background of states, i.e. the set of black holes of given mass, not all states in the Hilbert space. Secondly, we must consider the Fourier modes to be smeared to avoid becoming sensitive to the energy levels, which also introduces a time dependence.

\subsection{Lack of Left Inverse}
We now move to to another paradox \cite{Almheiri:2013hfa}.

\subsubsection{The Paradox} 
On one hand we know that the interior operator $\widetilde{b}_\omega^\dagger$ acts as a creation operator as we write, to leading order, the following
\begin{align}
\begin{split}
\bra{\Psi_0} \widetilde{b} \widetilde{b}^\dagger - \widetilde{b}^\dagger \widetilde{b} \ket{\Psi_0} &= \bra{\Psi_0} \widetilde{b} e^{\frac{\omega \beta} {2}} b - \widetilde{b}^\dagger e^{-\frac{\omega \beta} {2}} b^\dagger  \ket{\Psi_0} ,\\
&= \bra{\Psi_0} b b^\dagger  -  b^\dagger b  \ket{\Psi_0} ,\\
&= 1 ,
\end{split}
\end{align}
where we used that the interior and exterior operators commute and, secondly, we used $\widetilde{b} \ket{\Psi_0}=e^{-\beta H/2}b^\dagger e^{\beta H/2} \ket{\Psi_0}=e^{-\frac{\omega \beta} {2}} b^\dagger\ket{\Psi_0}$. We can, therefore, use $[\widetilde{b},\widetilde{b}^\dagger]=1$ inside correlators. Moreover, this should hold as an operator statement if the interior operators are state-independent, which means that $\widetilde{b}^\dagger$ must have a left-inverse.

On the other hand, we have that $[H, \widetilde{b}^\dagger]=-\omega \widetilde{b}^\dagger$ and, therefore, it lowers the energy of the state. There are fewer states at lower energies and $\widetilde{b}^\dagger$ must annihilate part of the typical state and, therefore, cannot have a left-inverse. We will see that this does not interfere with $C$, i.e. the kernel does not need to significantly overlap with $C$. 

\subsubsection{The Resolution}
We can avoid annihilating part of the state by fine-tuning the operators to the phases of the state, or by fine-tuning the phases of the state to the operators. This is equivalent to selecting a fine-tuned time of the state. We can check in a simple toy model whether mapping to a smaller set of states can leave the norm of vector intact. It is good to remember that $C$ is really small compared to the size of the Hilbert space. The phases are fixed this halves the dimensionallity of the space, and flipping the sign of an energy eigenstate is the same as time evolution. Only one combination of signs is in $C$, which is a $2^{-e^S}$ part of the real part of the Hilbert space.

We can use matrices as a simple toy model to test whether we can avoid annihilating part of the state by tuning the phase. For example, consider the following matrix equation:
$M \vec{y} = \vec{x}$ where $M$ is an $m \times n$ matrix, $\vec{y}$ is a length $n$ vector, and $x$ is a length $m$ vector. We consider the case where $m > n$ and compare the between $y$ with random complex elements and $\vec{y}$ with positive real elements. To be explicit, we pick $m=2$ and $n=1$ and consider $\vec{y}^\dagger M^\dagger M \vec{y}$, thus an example map that we are interested in is $M=\begin{bmatrix}1 & 0\end{bmatrix}$, which gives
\begin{equation}
M^\dagger M = \begin{bmatrix}
   1 & 0\\
   0 & 0
\end{bmatrix} .
\end{equation}
It is easy to see that this will partially annihilate the state and reduce the size of the vector to $\norm{\vec{x}}^2=\tfrac{1} {2}$, in both the case of the complex vector and of the `positive' vector. We can, however, change the basis and work with the similar matrix
\begin{equation}
M^\dagger M = \frac{1} {2} \begin{bmatrix}
   1 & 1\\
   1 & 1
\end{bmatrix} .
\end{equation}
Now we see that $\norm{\vec{x}}^2=(y_1^\dagger y_1 + y_2^\dagger y_1 + y_1^\dagger y_2 + y_2^\dagger y_2)/2$. For the complex factors the random phases of the cross terms average out and we are again left with $\norm{\vec{x}}^2=\tfrac{1} {2}$. This is different for the `positive' vector, because there are no phases and nothing cancels. This kind of map can be generalized to higher dimensions, where the elements of $\vec{y}$ become uncorrelated and the norm of the approaches $\norm{\vec{x}}^2=1 - \epsilon$, where $\epsilon$ is suppressed by the size of the vectors. In higher dimensions this generalizes to
\begin{equation}
\begin{bmatrix}
    1 & 0 & 0 & \dots  & 0 \\
    0 & 0 & 0 & \dots  & 0 \\
    \vdots & \vdots & \vdots & \ddots & \vdots \\
    0 & 0 & 0 & \dots  & 0
\end{bmatrix}
\rightarrow 
\frac{1} {m}
\begin{bmatrix}
    1 & 1 &1 & \dots  & 1 \\
    1 & 1 & 1 & \dots  & 1 \\
    \vdots & \vdots & \vdots & \ddots & \vdots \\
    1 & 1 & 1 & \dots  & 1
\end{bmatrix} .
\end{equation}
More generally, most entries of the matrix need to be positive, and that some of its eigenvalues are zero. It is clear that these maps reduce the size of the state, but they do not reduce the norm of the state (in the large $N$ limit).

We can use the same steps to obtain a matrix that reduces the norm of state. For example, we look at the following matrix
\begin{equation}
b=             
\begin{bmatrix}
   \sqrt{3}-1 & \sqrt{3}+1 & 2 &  0 \\
    1 & -1 & 1 &  3 \\
    \sqrt{3}+1 & \sqrt{3}-1 & 2  & 0
\end{bmatrix},
\end{equation}
such that
\begin{equation}
\frac{1}{12} b^\dagger b = \begin{bmatrix}
  1 & 0 & 0 &  0 \\
    0 & 1 & 0 &  0 \\
    0 & 0 & 1 &  0 \\
 0 & 0 & 0 & 1
\end{bmatrix} +\frac{1}{4}\begin{bmatrix}
   -1 & 1 & -1 &  1 \\
    1 & -1 & 1 &  -1 \\
    -1 & 1 & -1  & 1 \\
1 & -1 & 1 & -1
\end{bmatrix}.
\end{equation}
The alternating signs reduce the error we make when we use $\frac{1}{12}b^\dagger$ as an approximate left inverse provided that the elements of the vector we act on are positive. We can estimate the error by drawing random vectors $v$ with the Haar measure and calculate the error we make when we use $\frac{1}{12}b^\dagger$ as a left inverse.

\begin{table}[h]
\centering
\begin{tabular}{l|c}
                                   & $|1-v^\dagger (\frac{1}{12} b^\dagger b) v|$    \\ \hline
Free phases & 1.0783 \\
Fixed phases                           & 0.0538
\end{tabular}
\end{table}
where we averaged over $10^6$ random draws. The phases were fixed by taking the element wise absolute value of the random vector. We see that we cannot change the basis of the state without also changing the operators, while keeping the result the same, because changing the state would introduce phases. Thus we cannot change the basis of the state while working with interior operators, just as in section \ref{ssec:lipr}.

\subsection{Other Paradoxes}
There are several other paradoxes, which we will only discuss briefly as the approaches from the previous resolutions also applies to these.
\begin{enumerate}
  \item The Strong Subadditivity Puzzle.
\end{enumerate}
For a black hole to have a smooth horizon it is necessary that particles just outside the black hole are entangled with the particles just inside the horizon. Moreover, particles just outside the horizon must be entangled with early radiation for information to escape the black hole \cite{Almheiri:2013hfa}. However, monogamy of entanglement forbids this. This assumes that that the interior is independent of the exterior, which is something that we did not require, i.e. the interior can be constructed from fine-tuned multi-trace operators, which makes these operators dependent on the exterior.

\begin{enumerate}
  \setcounter{enumi}{1}
  \item The $[\text{Exterior} , \text{Interior}] \neq 0$ Paradox.
\end{enumerate}
Commutators between an operator and a scrambled version thereof tend to be order one \cite{Almheiri:2013hfa}.
\begin{equation}
[U{\cal O}U^\dagger,{\cal O}] \sim O(1),
\end{equation}
for a scrambling unitary $U$, which should be detectable outside the black hole. However, we do not require that the interior operators are of the form $U{\cal O}U^\dagger$ as we allow them to be fine-tuned, which gives the freedom for the interior and exterior operators to commute.

\section{The Natural Cone}
\label{sec:natcone}
There is a lot of freedom in choosing the set $C$, as we can replace each state with the time evolution of that state. The construction of Tomita and Takesaki gives a candidate for the set $C$, which does not have this ambiguity. The set of states of the form
\begin{equation}
{\cal P}_{\ket{\psi}} = \overline{\{AJAJ\ket{\psi} : A \in {\cal A}\}} ,
\end{equation}
is called the natural cone \cite{Bratteli:1979tw} and the states in this cone have the property that modular conjugation $J$ is the same for all cyclic states in the cone. The mirror operators constructed in section \ref{sec:fourier} are, therefore, the same for all states inside the natural cone. We will look at some other properties of the natural cone in the next subsection.

We propose to use $C \approx {\cal P}_{\ket{\psi}} \cap \{\text{Typical States}\}$, where all states come from the same energy window. The following algorithm can be used to obtain the mirror operators for most typical states from the same energy window.
\begin{enumerate}
  \item Pick a random typical state $\ket{\psi_0}$ to serve as a reference state.
  \item Construct the mirror operators, without timeshift (see section \ref{ssec:mirpos}), for $\ket{\psi_0}$.
  \item Construct the natural cone ${\cal P}_{\ket{\psi_0}}$.
  \item Find the time when a test state $\ket{\psi_1(t)}$ is close to the natural cone ${\cal P}_{\ket{\psi_0}}$, up to some $O(1/N)$ tolerance.
  \item Use the same mirror operators for $\ket{\psi_1(t)}$ and $\ket{\psi_0}$.
  \item The mirror operators for $\ket{\psi_1(0)}$ are obtained by evolving from $\ket{\psi_1(t)}$ and taking explicit time dependence into account.
  \item Precursors must be used to obtain the correct correlators, similar to equation \eqref{eqn:shift}.
\end{enumerate}
It is enough for the test state to come close to the natural cone in step 4, because a $O(1/N)$ perturbation of a state will not affect leading order results. Whether a state gets close to the natural cone is discussed in section \ref{ssec:passing}. In section \ref{sec:timedep}, we will discuss that this algorithm is independent of the choice of reference state at leading order.

\subsection{Basic Properties}
The modular conjugation $J$ and modular hamiltonian $\Delta$, also called the modular objects, are defined for a specific state, and may differ for different states. Modular theory, however, tells us that there are some relations between the modular objects of different states \cite{Bratteli:1979tw}. For example, if two states are related by a unitary $\ket{\phi} = U\ket{\psi}$ then the modular object for $\ket{\psi}$ and the algebra $U {\cal A} U^\dagger$ is $(U\Delta U^\dagger , U J U^\dagger)$.  

Another relation between modular objects can be found found when we look at the following set of states. 
\begin{equation}
{\cal P} = \overline{\{AJAJ\ket{\psi} : A \in {\cal A}\}} ,
\end{equation}
where ${\cal P}$ is called the natural positive cone associated with the pair $({\cal A}, \ket{\psi})$. The overline denotes the closure in the Hilbert space ${\cal H}$. Note that the states in this cone may not be normalized, but normalization does not remove any of its interesting properties. Some of the properties of ${\cal P}$ are
\begin{itemize}
\item ${\cal P} = \overline{\{\Delta^{-1/4}AA^\dagger \ket{\psi} : A \in {\cal A}\}}$ .
\item ${\cal P}$ forms a convex cone.
\item ${\cal P}$ is self-dual.
\item $\text{span}({\cal P}) = \text{span}({\cal H})$ .
\item $AJAJ{\cal P} \subset {\cal P}, \forall A \in {\cal A}$ .
\item $J \ket{\phi} = \ket{\phi}, \forall \ket{\phi} \in {\cal P}$ .
\item $\ket{\phi}$ is cyclic $\Leftrightarrow$ $\ket{\phi}$ is separating, $\forall \ket{\phi} \in {\cal P}$ .
\end{itemize}
Moreover, if $\ket{\phi} \in {\cal P}$ is cyclic (and, therefore, separating) then the modular conjugation is the same $J_{| \phi \rangle} = J$ and the natural positive cone is the same ${\cal P}_{| \phi \rangle} = {\cal P}$.

The fact that ${\cal P}$ is self-dual means that the dual of ${\cal P}$, defined by
\begin{equation}
{\cal P}^D = \{ \ket{\psi} \in {\cal H} : \braket{\psi | \phi} \geq 0, \forall \ket{\phi} \in {\cal P}  \} ,
\end{equation}
is equal to ${\cal P}$. 

We proposed to use $C \approx {\cal P} \cap \{\text{Typical States}\}$, where all states come from the same energy window. The main advantage is that the modular operator $J$ is the same in the natural cone and the mirror operators are, therefore, state-independent within the natural cone. It is, thus, important that every typical state gets close to some element from the natural at some point in time,
\begin{equation}
\label{eqn:natconlap}
\ket{\Psi_1(t_0)} + \ket{\delta} = a_0 \ket{\Psi_0} + a_1 A_1 \widetilde{A}_1 \ket{\Psi_0} + a_2 A_2 \widetilde{A}_2 \ket{\Psi_0}...
\end{equation}
where $a_i>0$,  and the norm of $\ket{\delta}$ is small compared to the norm of $\ket{\Psi_1(t_0)}$. The operators $A_i$ that are most useful are of a specific form, see equation \eqref{eqn:basop}, which we will discuss later. The motivation that this superposition results in a small error term comes from the property that the overlap between different typical states is significant \cite{Shenker:2013yza}. For example 
\begin{equation}
\max_t (\langle \Psi_1(t) | \Psi_0 \rangle )= \frac{\pi} {4} ,
\end{equation}
which will only increase when we add more degrees of freedom, see appendix \ref{app:overlap} for more examples.

\subsubsection{Restricting to the Small Algebra}
\label{ssec:SAlg}
Most properties are conserved when we restrict to the energy window. However, it is only self-dual if we discount the part of the dual cone outside the energy window. 

The second restriction we must make is restricting to the small algebra. The main property we must test is whether the restricted cone spans the energy window. To do that we estimate how much volume the new cone fills. The easiest way to do this is to count the number of degrees of freedom we have for the construction of the natural cone, while we restrict ourselves with operators $AJAJ$ that stay within the energy window.

A single Fourier mode works as $[H,{\cal O}_\omega \tO_\omega]=0$. This, however, is badly behaved as a state, ${\cal O}_\omega \tO_\omega \ket{\Psi}$, for example, is not normalizable. We must smear the Fourier mode to remedy this problem. The width of the frequencies that we smear over, $\omega_s$, must be very small as the smearing causes ${\cal O}_\omega \tO_\omega \ket{\Psi}$ to be slightly wider than the energy window. We can view this spillover, or the removal of the spillover, as a perturbation as long as it is small. On the other hand, the smearing cannot be too small, because it should be at least a few level spacings in size, thus $O(e^{-S}) < \omega_s < O(1/N)$. 

Moreover, the frequency of the Fourier mode used cannot be too large. Less than some cut-off frequency $\omega_*$, as discussed in \ref{sec:fourier}. There is some discussion on how large this frequency can be, but we will be conservative in the following estimates.

We can multiply Fourier modes to obtain another operator that fits our requirements. So we can use the following operator
\begin{equation}
\label{eqn:basop}
A={\cal O}_{\omega_1}^{n_1}{\cal O}_{\omega_2}^{n_2}{\cal O}_{\omega_3}^{n_3}...
\end{equation}
where we multiplied the operator with frequency $\omega_i$ times itself $n_i$ times. All $N_i$ must be smaller than some $n_m = O(N^0)$, as the states we are looking for are almost in thermal equilibrium and we should not deviate too far from that. We assume that the different frequencies are distinct as we binned them with $\omega_s$, which means that the ordering does not matter. In appendix \ref{app:thermal}, we show that that states perturbed by single Fourier modes in the discussed manner are independent from each other, even when the frequencies are almost the same.

The number of states obtained is given by the number of different operators we can construct, as detailed above.
\begin{equation}
\#\text{States} = (n_m+1)^{(n_p \omega_* / \omega_s)} ,
\end{equation}
where $n_p=O(N^0)$ is the number light primaries in the theory. The plus one comes from the possibility that a Fourier mode is absent in the operator. This estimate is exponentially large and certainly comparable with $2^n$ , but it is an overcounting as we have not restricted sum of frequencies to be below the frequency cut-off. This, however, does not change the order of magnitude, as the size of the smearing is a much larger contribution to the number of operators that we can use to construct new states.

These counting arguments are reminiscent of the counting of 't Hooft brick wall model \cite{'tHooft:1984re}, where the correct entropy can be derived from assuming a cut-off provided by a brick wall at the horizon. 

We will assume that we can use the properties of the previous subsection, even when using the small algebra. This is, however, the main weakness of this proposal, as there is direct competition between the number of states that we can create in this manner, which we want to be large, and the size of the small Hilbert space, which we want to be small.

\subsubsection{Going from any State to ${\cal P}$}
\label{ssec:passing}
\begin{figure}
\centering
    \includegraphics[width=.6\textwidth]{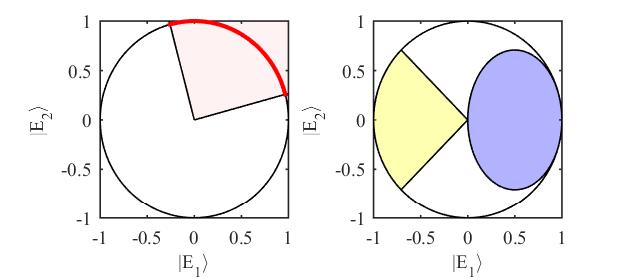}
\caption{Left) We start with $\mathbb{C}^2$ and choose two random orthogonal vectors in this space. This spans a real subspace $\mathbb{R}^2$, and positive superpositions of the two vectors form a self-dual cone. The red circle arc is the unit normalized part of the self-dual cone. \quad Right) The three dimensional analogue, where we show the upper hemisphere and the self-dual orthant (yellow) and circular (blue) cones.  }
\label{fig:coneexm}
\end{figure}

The natural cone ${\cal P}$ is a self-dual cone. This means that the inner product between two states, $\ket{\Psi},\ket{\Phi}$ from the natural cone is positive $\langle \Phi | \Psi \rangle > 0$. It is, however, useful to consider the real subspace ${\cal H}_{\cal P}$ that contains the natural cone ${\cal P}$. This means that the inner product between any state $\ket{\chi}$ from the real subspace ${\cal H}_{\cal P}$ and any state from the natural cone $\ket{\Psi}$ is real, $\langle \chi | \Psi \rangle \in \mathbb{R}$. We can do this because both ${\cal P}$ is convex, and spans the small Hilbert space. We can, therefore, write any vector in $\ket{\chi} \in {\cal H}_{\cal P}$ as a sum over vectors $\sum_i c_i \ket{\psi_i}$, where $\ket{\psi_i} \in {\cal P}$ and $c_i \in \mathbb{R}$. It is useful to consider these real subspaces as it is much more intuitive what the cones look like in a real subspace. We show a low dimensional toy example in figure \ref{fig:coneexm}. 

If we make some simplifying assumptions, we can make an estimate whether a typical state $\ket{\Psi}$ passes through ${\cal P}$ under time evolution.

\begin{enumerate}
  \item $\ket{\Psi}$ passes through ${\cal H}_{\cal P}$ at independent uniformly random points.
\end{enumerate}
We allow an exponentially small tolerance when we consider whether a state passes through ${\cal H}_{\cal P}$ at some time. This is necessary because there are, generally speaking, energy levels that are not related by rational numbers. Therefore, it would take an infinite amount of time get a specific phase alignment between the energy levels. 

The assumption that the points are uniformly distributed is too strong, as correlations that do exist tend to push the points at which the state passes through ${\cal H}_{\cal P}$ away from each other. This happens because the inner product $\bra{\Psi} e^{-iHt} \ket{\Psi}$ goes to zero very fast, meaning that at large time separation the states are almost orthogonal.

This means that at each pass through ${\cal H}_{\cal P}$ the state has a probability $P_1$ to be in ${\cal P}$, which is given by the ratio of the volumes
\begin{equation}
\label{eqn:p1}
P_1=\frac{\text{Vol}({\cal P}\cap E_\delta \cap S^{n-1})} {\text{Vol}({\cal H}_{\cal P} \cap E_\delta \cap S^{n-1})} \geq 2^{-n},
\end{equation}
where $E_\delta$ is the energy window, $n$ is the number of states in the energy window, and $ S^{n-1}$ is the $(n-1)$-sphere to force the states to be normalized. See appendix \ref{app:cones} for more discussion about the volume of self-dual cones. Here we have used the property that ${\cal P}$ is dense and spans the Hilbert space, which also defines the size of ${\cal H}_{\cal P}$.

\begin{enumerate}
  \setcounter{enumi}{1}
  \item $\ket{\Psi}$ passes through ${\cal H}_{\cal P}$ often, namely $2^n$ times,
\end{enumerate}
where $n$ is the number of states in the energy window. Again, we use an exponentially small tolerance, thus points that differ by an amount less than the tolerance are identified. This is easy to prove for the real subspace spanned by the energy eigenvectors. A typical state passes through this subspace when all magnitudes are real, either positive or negative. We, therefore, pass $2^n$ times through this subspace, hitting all combinations of signs of the energy eigenvectors. It is reasonable to assume that any state passes $2^n$ times through ${\cal H}_{\cal P}$ as well, because they same degrees of freedom.

Another way to see this, is to look at equation \eqref{eqn:timestate} and see that, for a reasonable distribution of $a_{ij}$, the state passes through ${\cal H}_{\cal P}$ at least the assumed number of times.

Combining this with the result of the last assumption allows us to estimate the probability that any typical state passes through the natural cone
\begin{equation}
P_2 = 1-(1-P_1)^{(2^n)}\simeq1-e^{-1} \approx 0.6321 .
\end{equation}
This rough estimate of the probability is an $O(N^0)$ number, but is significantly smaller then one. We are, therefore, not certain that all states pass through the natural cone.

This is a very conservative estimate. For example, looking at equation \eqref{eqn:timestate} the number of passes through ${\cal H}_{\cal P}$ may be much larger then $2^n$, in which case $P_2$ goes to one. Moreover, if the natural cone is more rounded, see appendix \ref{app:cones}, than the ratio of volumes is given by $a^{-n}$, with $\sqrt{2} < a < 2$, in which case $P_2$ goes to one as well. We can effectively get $P_2=1$ with one additional assumption, even with this conservative estimate.

\begin{enumerate}
  \setcounter{enumi}{2}
  \item Small perturbation around the state $\ket{\Psi}$ cause large changes in the time it takes to pass through ${\cal H}_{\cal P}$.
\end{enumerate}
The phases need to be extremely fine tuned for a state to be in any specific real subspace. Thus while one state is in ${\cal H}_{\cal P}$, a different state that started close to the original state may need a lot of time before it passes through ${\cal H}_{\cal P}$ as well. Because of the large time differences a perturbed states passes through ${\cal H}_{\cal P}$ at points independent of the original state. 

For example, consider a state with all coefficients real and positive at $t=0$. It is, therefore, in the real subspace spanned by the energy eigenstates with real coefficients at time  $t=0$. Now consider a perturbation of this state, which gives a complex phase to a single energy eigenstate. This causes a large timeshift in the times when the perturbed state passes through the real subspace spanned by the energy eigenstates with real magnitudes. The points at which the perturbed states passes through the subspace is the same as the original state. This, however, is a consequence of the property that the energy eigenvalues form a real basis for the subspace and are the eigenstates of the Hamiltonian. This is not the case for ${\cal H}_{\cal P}$, and time evolution of a state will, therefore, change the magnitudes of the real basis vectors. We can write the time evolution of a typical state in the basis of the real subspace ${\cal H}_{\cal P}$ as follows
\begin{equation}
\label{eqn:timestate}
\ket{\Psi(t)}=\sum_i c_i e^{- i E_i t+\theta_i} \ket{E_i} = \sum_{i,j} a_{ij} e^{-i E_i t+\theta_{ij}}\ket{j} ,
\end{equation}
where we selected the coefficients $c_i, a_{ij}$ to be real and positive and explicitly wrote the phases $\theta_i,\theta_{ij}$. The state lies on ${\cal H}_{\cal P}$ when the coefficient $\sum_{i} a_{ij} e^{-i E_i t+\theta_{ij}}$  in front of the basis vectors $\ket{j}$ is real for all basis vectors.

\begin{figure}
\centering
    \includegraphics[width=.6\textwidth]{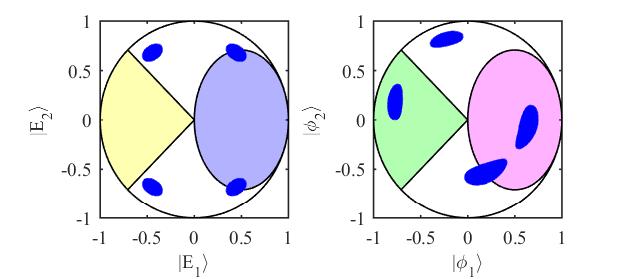}
\caption{Left) Upper hemisphere is shown with the orthant (yellow) and circular (blue) cones. The blue dots are where a random test vector with time evolution gets close to the $\mathbb{R}^3$  subspace of $\mathbb{C}^3$, which is spanned by the energy eigenvalues. \quad Right) The subspace is spanned by random vectors which causes the places where the test vector gets close the real subspace to shift and stretch. This shows that it is important that the natural cone is not contained in the real span of the energy eigenvectors.}
\label{fig:hitexm}
\end{figure}

We can almost always find a state $\ket{\Psi_1}$ close to the original state $\ket{\Psi_0}= \ket{\Psi_1} +\frac{1} {N} \ket{\delta}$ that does pass through ${\cal P}$.  The mirror operators $\tO = J{\cal O}J$ are bounded operators, the small perturbation from $\ket{\delta}$ will, therefore, only lead to subleading corrections. We can, therefore, use the mirror operators as fixed operators.

\subsubsection{Some Subtleties}
These results do not mean that we can apply the Tomita-Takesaki construction once and be done for all states. The requirement that the state is cyclic can exclude some states, but should be approximately true for most (almost all) typical states. It may seem that the restriction to the energy window is unnecessary, but this is not the case.

When applied to the the black hole typical states we want the equation to be true at leading order, and this restricts the algebra ${\cal A}$. Most typical states cannot be annihilated by a small number of operators, small compared to the energy of the black hole. We, therefore, have a dependence on the energy window in the algebra.

Moreover, the modular Hamiltonian is not necessarily the same for all states in ${\cal P}$. Physics, however, tells us that we should identify the modular Hamiltonian as in equation \eqref{modularh} for most typical states. We again have a dependence on the energy window, in the form of the inverse temperature $\beta$. It can be expected that the various approximations cause subleading corrections, which makes correlators with both $\tO$ and ${\cal O}$ more sensitive, at the subleading level, to the microstate. 

Time evolution does not change the cyclic and separating properties of states. Moreover, modular conjugation $J$ is the same for all cyclic and separating states in ${\cal P}$. We can, thus, consider $\tO=JOJ$ to be a fixed operator when working on any typical state up to time evolution, as any state can almost be mapped into ${\cal P}$ by time evolution for a state-dependent amount. We will discuss the time-dependence in section \ref{sec:timedep}.

Finally, there will be states that have strange time evolution compared to the typical states. Energy eigenstates only change by an overall phase and may be problematic for this construction. However, such states only constitute a small number of states in the energy window and are, therefore, not of interest to us. These caveats do not alter the conclusion that almost all state get close to ${\cal P}$ under time evolution.

\subsection{Application of the Fixed Mirror Operators}
\label{sec:appl}
In this subsection, we discuss how we can use the mirror operators in correlators.

\subsubsection{How to Apply}
Let us reiterate the rules we used when we applied the mirror operators in correlation functions.
\begin{align}
\label{eqn:mirrules}
\begin{split}
\widetilde{{\cal O}} \ket{\Psi_0} =& \; e^{-\frac{\beta H} {2}} \mathcal{O}^\dagger e^{\frac{\beta H} {2}} \ket{\Psi_0} , \\
\tO A \ket{\Psi_0} =& \; A \tO\ket{\Psi_0} , \\
[H,\tO] A \ket{\Psi_0} =& \;  A e^{-\frac{\beta H} {2}}[H, \mathcal{O}^\dagger] e^{\frac{\beta H} {2}} \ket{\Psi_0} ,
\end{split}
\end{align}
where we left out the time arguments, as we will discuss them in the next subsection. These rules must be the same to leading order when we use a different typical state in the natural cone as reference state, for example,
\begin{equation}
\ket{\Psi_1} = A\widetilde{A}\ket{\Psi_0} .
\end{equation}
The second is satisfied for $\ket{\Psi_1}$, because the modular conjugation $J$ is the same in the natural cone. The third line is satisfied if the first line holds, because we identified the modular Hamiltonian for typical states as in equation \eqref{modularh}. So we only need to check the consistency of the first line. Using $\ket{\Psi_0}$ as the reference state we have
\begin{equation}
\label{eqn:comp}
  \bra{\Psi_1} B_1 \widetilde{B}_2 \ket{\Psi_1} = \bra{\Psi_0} A^\dagger B_1 A e^{-\frac{\beta H} {2}} (A^\dagger B_2 A)^\dagger e^{\frac{\beta H} {2}}  \ket{\Psi_0},
\end{equation}
while consistency of the first line of \eqref{eqn:mirrules} demands that this is close to using $\ket{\Psi_1} = A\widetilde{A}\ket{\Psi_0}$ as the reference state
\begin{equation}
\label{eqn:comp2}
\bra{\Psi_1} B_1 \widetilde{B}_2 \ket{\Psi_1} = \bra{\Psi_1} B_1 e^{-\frac{\beta H} {2}} B_2^\dagger e^{\frac{\beta H} {2}} \ket{\Psi_1} =\bra{\Psi_0} A^\dagger B_1 e^{-\frac{\beta H} {2}} B_2^\dagger e^{\frac{\beta H} {2}} A e^{-\frac{\beta H} {2}} A^\dagger  A e^{\frac{\beta H} {2}} \ket{\Psi_0} .
\end{equation}
These two are, certainly, not equal for all $A$. However, $A$ is restricted for typical states as discussed in section \ref{ssec:SAlg}.

\subsubsection*{A Simple Example}
We are interested in the case where $A$ is constructed from Fourier modes with narrow spread, as shown in equation \eqref{eqn:basop}. We will work out what happens for a single Fourier mode and generalize from there. It is useful to work with the following modes
\begin{align}
\begin{split}
\langle a^\dagger a \rangle &= \frac{1} {e^{\beta \omega} - 1} \delta (\omega-\omega') = g(\omega) \delta(\omega-\omega') ,\\
\langle a a^\dagger \rangle &= \frac{e^{\beta \omega}} {e^{\beta \omega} - 1} \delta (\omega-\omega') =e^{\beta \omega} g(\omega) \delta(\omega-\omega')=g(-\omega) \delta(\omega-\omega') ,\\
\end{split}
\end{align}
and consider $A=a$ up to normalization, which we can work out as follows
\begin{equation}
\bra{\Psi_0}a^\dagger \widetilde{a}^\dagger \widetilde{a} a \ket{\Psi_0} = g(\omega)(1+2g(\omega))\delta^2 ,
\end{equation}
where we omitted the arguments of the delta functions. Detailed calculations can be found in appendix \ref{app:thermal}. The delta functions disappear when we integrate over the smearing of Fourier modes. We have to integrate four times, while we have two delta functions. We, therefore, obtain an expression for the normalization of the state $\ket{\Psi_1} = a\widetilde{a}/c_1 \ket{\Psi_0}$
\begin{equation}
c_1 = \omega_s \sqrt{g(\omega)(1+2g(\omega))} ,
\end{equation}
where $\omega_s$ is the width of the smearing. We can now try to compare equations \eqref{eqn:comp} and \eqref{eqn:comp2} with this preliminary work done. For this example we will use that the operator $B_2$ is some integral over Fourier modes $\int_0^{\omega^*}d\omega_2K(\omega_2)b_{\omega_2}$.
\begin{align}
\begin{split}
 \bra{\Psi_0} A^\dagger B_1 A e^{-\frac{\beta H} {2}} (A^\dagger B_2 A)^\dagger e^{\frac{\beta H} {2}}  \ket{\Psi_0} &=\int_0^{\omega^*}d\omega_2K(\omega_2) c_1^{-2} \bra{\Psi_0} a^\dagger B_1 a e^{-\frac{\beta H} {2}} (a^\dagger b_{\omega_2} a)^\dagger e^{\frac{\beta H} {2}}  \ket{\Psi_0}\\
&=\int_0^{\omega^*}d\omega_2K(\omega_2) c_1^{-2}e^{\frac{\omega_2 \beta} {2}} \bra{\Psi_0} a^\dagger B_1 a  a^\dagger   b_{\omega_2}^\dagger  a)   \ket{\Psi_0} ,
\end{split}
\end{align}
while the second line is
\begin{align}
\begin{split}
&\bra{\Psi_0} A^\dagger B_1 e^{-\frac{\beta H} {2}} B_2^\dagger e^{\frac{\beta H} {2}} A e^{-\frac{\beta H} {2}} A^\dagger  A e^{\frac{\beta H} {2}} \ket{\Psi_0}\\ &\quad= \int_0^{\omega^*}d\omega_2K(\omega_2)c_1^{-2}\bra{\Psi_0} a^\dagger B_1 e^{-\frac{\beta H} {2}} b_{\omega_2}^\dagger e^{\frac{\beta H} {2}} a e^{-\frac{\beta H} {2}} a^\dagger  a e^{\frac{\beta H} {2}} \ket{\Psi_0}\\
&\quad= \int_0^{\omega^*}d\omega_2K(\omega_2)c_1^{-2}e^{\frac{\omega_2 \beta} {2}} \bra{\Psi_0} a^\dagger B_1  b_{\omega_2}^\dagger  a  a^\dagger  a  \ket{\Psi_0}.
\end{split}
\end{align}
We can, therefore, write the error we make, when we use the second line of equation \eqref{eqn:comp} instead of the first line, as
\begin{align}
\begin{split}
\text{error} &=\int_0^{\omega^*}d\omega_2K(\omega_2) c_1^{-2}e^{\frac{\omega_2 \beta} {2}} \bra{\Psi_0} a^\dagger B_1  [b_{\omega_2}^\dagger,  a  a^\dagger]  a  \ket{\Psi_0}\\
&= -\int_0^{\omega^*}d\omega_2K(\omega_2)c_1^{-2}e^{\frac{\omega_2 \beta} {2}} \bra{\Psi_0} a^\dagger B_1 a^\dagger a  \ket{\Psi_0} \delta(\omega-\omega_2),
\end{split}
\end{align}
where we assumed that $b$ and $a$ are Fourier modes of the same operator and, therefore, obey canonical commutation relations. The constants $c_1$ in front are not important as we need to compare the error term to the size of equation \eqref{eqn:comp} and that will have the same factor. The delta function limits the integration over the correlator with the integration kernel to the frequency window given by $\omega_s$. Assuming that the kernel is smooth at $\omega$, we see that the error term has a factor of $K(\omega)\omega_s$ relative to the correlators that we are comparing the error to.

We see that $\ket{\Psi_1}$ is no longer a typical state, as the state does not have a thermal distribution for the mode around $\omega$. This does not change expectation values for operators in position space, because they are not sensitive to disturbances to single Fourier modes. Moreover, summing over many different disruptions, as in equation \eqref{eqn:natconlap}, should average out and return to an approximately thermal state, i.e. a typical state.

\subsubsection*{Generalization}
We can generalize the previous example by considering $A=b_{\omega_1}b_{\omega_2}b_{\omega_3}...$. Each frequency will generate a term with a delta function at that frequency. The sum over these cannot grow too large as the phases are usually random. The sum, therefore, only scales as $\sqrt{n}$, where $n$ is the number of operators used. The maximum number of operators that we can use is also limited, as discussed in section \ref{ssec:SAlg}, which ensures that the error will remain small.

The second generalization to consider are positive superpositions of these states. The same argument applies here. When we use a superposition of $n$ states, we would have $n^2$ correlators. The phases, however, are random and the sum of these correlators only scale as $n$, which cancels with the normalization.

Working with more complicated operators $B_1,B_2$ also increases the number of error terms. We, however, consider simple operators, which have a bulk interpretation, and these operators are simple enough not to cause problems. The growth of the error is expected as more complicated operators are more sensitive to $O(1/N)$ corrections.

Perturbing the state a little bit with $\ket{\delta}$ introduces more error terms, see equation \eqref{eqn:natconlap}. The norm of this perturbation is, however, limited in size, which keeps these error terms under control.

These generalizations are enough for the cases that we are interested in. It is important to note that we clearly see subleading correction in the case that we have both normal and mirror operators in a correlator, while in the other cases the corrections are exponentially small, see section \ref{sec:lloyds}. We conclude that we can apply equation \eqref{eqn:mirrules} also to $\ket{\Psi_1}$ at leading order, while the subleading corrections are different.

\subsubsection{Time-Dependence of the Mirror Operators}
\label{sec:timedep}
 We discussed the mirror operators in position space in section \ref{ssec:mirpos}. We recall two of the equations that are useful for the discussion at this point. The Fourier transform
\be
\tO_T(t) = \int_{-\omega_*}^{\omega_*} d\omega\,\, e^{-i \omega (t-T)}  \tO_\omega ,
\ee
shows two key aspects of the explicit time-dependence. First, the sign in front of the physical time $t$ has the opposite sign from what we are used to in the Fourier transform. This means that the mirror operators go against the Killing isometry in the bulk. Secondly, there is the freedom of choosing $T$, which determines how much time evolution the operator has relative to the physical time. In position space we can approximately write 
\begin{equation}
\widetilde{{\cal O}}_T (t) \ket{\Psi_0} = e^{-\frac{\beta H} {2}} \mathcal{O}^\dagger(T-t) e^{\frac{\beta H} {2}} \ket{\Psi_0} ,
\end{equation}
where we clearly see the sign flip of $t$ and the time-shift by $T$. The other application rules deal with commutation relations, which are not affected by time dependence. This is enough to detail the time dependence of the operators when the mirror operators are defined for a single state.

We need to be more careful to get a consistent picture when we try to define the mirror operators for most typical states. It is natural to set $T=0$ for the reference state $\ket{\Psi_0}$. This also sets $T=0$ for typical states in the natural cone ${\cal P}$. For other typical states we use equation \eqref{eqn:natconlap} to find the amount of time evolution that we need to get close to the natural cone, and use $T$ to compensate that time evolution.

When doing an experiment with a typical state at some later time $\ket{\Psi_1 (t)}$, we can always recover which mirror operator is highly entangled with an exterior operator by setting $T=2t$ and work with the precursor of the mirror operator for that experiment. For example,
\begin{equation}
\bra{\Psi_1(t_0)} \tO_{2t_0} (t_0+t') \mathcal{O}(t_0+t'') \ket{\Psi_1(t_0)} = \bra{\Psi_1(t_0)} \mathcal{O}(t_0+t'') e^{-\frac{\beta H} {2}} \mathcal{O}^\dagger(t_0-t') e^{\frac{\beta H} {2}} \ket{\Psi_1(t_0)} ,
\end{equation}
this correlator is a function of $(t''+t')$ and is independent of $t_0$, and shows the entanglement structure at time $t_0$. The freedom to use the precursors is enhanced, because there is no reason to assume that the complexity of the mirror operators is lower at specific times.

It is important that we use a consistent time-shift $T$ when we work with the precursors. We lose a consistent bulk interpretation otherwise. It is worth noting that all state-dependence is captured in the selection of $T$, i.e. if we want to construct a mirror operator with maximal entanglement at that time for a specific state, we find $T$ and use it with the mirror operators from the reference state to generate the mirror operators for that specific state.

The modular conjugation operator $J$ is approximately constant in time. This can be seen for the following property $J\Delta J = \Delta^{-1}$. This together with the antilinear nature of $J$ means that $J\Delta^{is} J = \Delta^{is}$. However, time evolution is only approximately equal to $\Delta^{is}$, which introduces a small amount of time evolution in the modular conjugation operator $J$ and, therefore, in the definition of the Fourier modes of the mirror operator. Time evolution of $J$ also introduces a time dependent phase. This phase, however, is not relevant as the correlation functions we consider always use an even number of $J$'s.

\subsubsection*{Choice of Reference State}
Most of the discussion so far made use of a reference state $\ket{\Psi_0}$ in their definitions, initially for the mirror operators and later for the natural cone ${\cal P}$. This, however, seems contrary to the claim that this construction is state-independent, except for the explicit time dependence. We already know that the mirror operators are the same for a typical state from the natural cone of the reference state, as $\ket{\Psi_1} \in {\cal P}_{\ket{\Psi_0}} \Rightarrow {\cal P}_{\ket{\Psi_1}} ={\cal P}_{\ket{\Psi_0}}$, provided that $\ket{\Psi_1}$ is cyclic.

We limit the discussion to the moment in time that a typical state is close to the natural cone for other typical states, as explicit time evolution will generate the answers for the other states. The leading results are the same for the reference state and the other typical state. Subleading results, however, may differ. The subleading corrections are, therefore, relative to the reference state and will change when we select a different reference state.

Another object that depends on the reference state is the small Hilbert space. The natural cone is part of the small Hilbert space, ${\cal P}_{\ket{\Psi_0}} \subset {\cal H}_{\ket{\Psi_0}}$, even though the product $A\widetilde{A}$ is not part of the small algebra. This happens because we can convert the mirror into normal operators as it is next to the state.

For any other typical state, at the moment that they are close to the natural cone, we have
\begin{equation}
{\cal H}_{(\ket{\Psi_1}+\ket{\delta})} = {\cal H}_{\ket{\Psi_0}} ,
\end{equation}
moreover, time evolution does not alter the Hilbert space structure, so this holds at all times.

\subsection{Consistency Checks}
The construction is consistent when comparing different typical states. There are, however, other consistent checks that we need to do.

\subsubsection{Superpositions}
Superpositions are fundamental in quantum mechanics and our construction should be consistent when taking superpositions. 

We can always take positive superpositions in the natural cone and remain in the natural cone because of the conical structure. This extends to superpositions of typical states that are at the same time, if we have defined $t=0$ as the moment that they are close to the natural cone. 

Other superposition, either not strictly positive or of typical states at different times, are problematic. These superpositions can result in a state that has a different time-shift compared to the states it is a superposition of. This means that the bulk interpretation the superposition can be very different from the bulk interpretations of the states that made up the superposition. This is different from the exterior, where we can interpret the superposition of states that have a particle in the exterior ${\cal O}(x,t)c_1\ket{\Psi_1}+{\cal O}(x,t)c_2\ket{\Psi_2}={\cal O}(x,t)\ket{\Psi_3}$ as being a state having a particle in the same position. The interior is dependent on the time-shift and when the particle appears is not necessarily at the same time as before the superposition.

It is also possible to construct states that never get close to the natural cone and the interior operators cannot be defined for those states. This may seem rather destructive for the construction, but it is exactly what we expect. We can create any state by taking arbitrary superpositions, including states with aberrant behavior. It is, therefore, consistent that we cannot take arbitrary superpositions, while keeping the interior operators fixed.

\subsubsection{Perturbations}
A critique of the original mirror operators \cite{Harlow:2014yoa} was that a perturbed state of the form $\ket{\Psi_1}=e^{ig \tO_\omega\tO_\omega^\dagger}\ket{\Psi_0}$ is also an equilibrium state and that we can define mirror operators for both $\ket{\Psi_1}$ and $\ket{\Psi_0}$, and conclude that both have smooth horizons. That both states have smooth horizons is in itself is not inconsistent. We must be careful when look at a state of the following form
\begin{equation}
\ket{\Psi(t)} = \begin{cases} \ket{\Psi_0(t)} & t<0 , \\ \ket{\Psi_1(t)} & t>0 . \end{cases}
\end{equation}
Here we notice that it takes a time-shift to get  $\ket{\Psi_1}$ close to the natural cone, as $e^{ig \tO_\omega\tO_\omega^\dagger}\ket{\Psi_0}$ is not in the natural cone. Explicit time dependence reduces correlation between mirror operators and normal operators after the perturbation, with a firewall as a consequence as expected by \cite{Harlow:2014yoa}.

\section{Conclusions}
In this paper we discussed whether explicit time dependence is enough to avoid the firewall paradoxes. This was inspired by the state-dependent mirror operators and the property that ergodic systems pass through all states of a given energy under time evolution. We have shown that it is possible to avoid the paradoxes with explicit time dependence, even though typical black hole microstates do not pass through all other typical black hole microstates under time evolution.

We have proposed to use the natural cone to get a construction of the interior operators. We have provided evidence that the natural cone defines the modular operator $J$ for most typical states, thereby reducing the state dependence to a single parameter. Moreover, we know what this parameter does as this parameter comes from time evolution. This reduced state-dependence, therefore, reduces the fine-tuning needed for a smooth horizon and allows easier explicit construction in toy models.

\begin{figure}[t]
\centering
    \includegraphics[width=.6\textwidth]{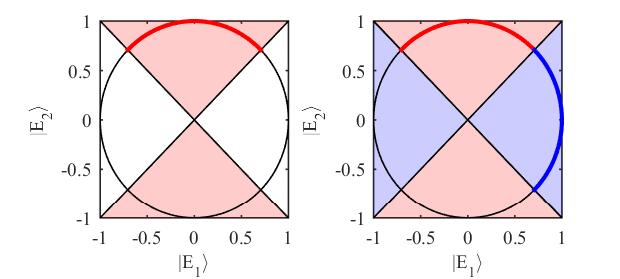}
\caption{Left) In $\mathbb{C}^2$ it is possible to select a cone that only captures half of the states with time evolution. \quad Right) We add a second second cone to fill the space and are now able to capture all states.}
\label{fig:multexm}
\end{figure}

Alternatively, if the typical states are not all captured in the natural cone and the time evolution thereof, then the typical states can be decomposed in multiple natural cones, see figure \ref{fig:multexm} for a toy model example. Remember that the natural cones of different states are the same if one state is in the natural cone of the other. Thus, if a test state and its time evolution is not in the natural cone of a reference state, then the natural cone of the test state has no overlap with the natural cone of the reference state. We can, therefore, define the mirror operators piece wise on independent natural cones, provided that the number of independent natural cones is not too large.

\section{Acknowledgments}
The author would like to thank Jan de Boer, Suvrat Raju, and Erik Verlinde for discussions. In particular, the author would like to thank Kyriakos Papadodimas for extensive discussions and comments on the draft. The author would like to thank NCCR SwissMAP of the Swiss National Science Foundation for financial support.

\appendix

\section{Thermal Correlators}
\label{app:thermal}
In this appendix we investigate several correlators that are relevant to the discussions in section \ref{ssec:SAlg} and section \ref{sec:appl}. We will only focus on the leading order in all calculations in this appendix.

For a Fourier mode $a$ at frequency $\omega$  we can write the commutation relations as
\begin{equation}
[a,a^\dagger]=\delta(\omega-\omega') .
\end{equation}
Moreover, we can use the commutation relations with the Hamiltonian to obtain the following expression
\begin{equation}
e^{-\beta H} a = e^{\beta \omega} a e^{-\beta H} .
\end{equation}
These two equations are enough to calculate the correlators we are interested in.

\begin{align}
\begin{split}
\langle a^\dagger a \rangle &= \Tr[e^{-\beta H} a^\dagger a]/Z = -\delta(\omega-\omega')+ \Tr[e^{-\beta H} a a^\dagger]/Z = -\delta(\omega-\omega')+ e^{\beta \omega}\Tr[e^{-\beta H} a^\dagger a]/Z \\
&= \frac{1} {e^{\beta \omega}-1} \delta(\omega-\omega') ,
\end{split}
\end{align}
where we used that the correlators in the micro-state are thermal to leading order. We obtain among similar lines the following
\begin{equation}
\langle a a^\dagger \rangle  =\frac{e^{\beta \omega}} {e^{\beta \omega}-1} \delta(\omega-\omega') .
\end{equation}

We can also obtain the two point function between a normal and a mirror operator in this manner
\begin{equation}
\langle a \widetilde{a} \rangle = \langle a e^{-\beta H/2}a^\dagger e^{\beta H/2} \rangle =e^{-\beta \omega/2} \langle a a^\dagger \rangle = \frac{e^{\beta \omega/2}} {e^{\beta \omega}-1} \delta(\omega-\omega') ,
\end{equation}
and similarly
\begin{equation}
\langle a^\dagger \widetilde{a}^\dagger \rangle = \frac{e^{\beta \omega/2}} {e^{\beta \omega}-1} \delta(\omega-\omega') .
\end{equation}

We are also interested in the norm of a state of the form $\ket{\Psi_1} \sim \widetilde{a}a\ket{\Psi_0}$, where the Fourier modes are slightly smeared. We, therefore, calculate the following four point function
\begin{align}
\begin{split}
\langle a^\dagger \widetilde{a}^\dagger \widetilde{a} a \rangle &= \langle a^\dagger  a a^\dagger a \rangle \\
&=\Tr[e^{-\beta H} a^\dagger a a^\dagger a]/Z \\
&=\Tr[e^{-\beta H} a a^\dagger a^\dagger a]/Z - \Tr[e^{-\beta H} a^\dagger a]/Z \delta(\omega-\omega')\\
&=\Tr[e^{-\beta H} a a^\dagger a a^\dagger]/Z - \Tr[e^{-\beta H} a^\dagger a]/Z \delta(\omega-\omega') - \Tr[e^{-\beta H} a a^\dagger]/Z \delta(\omega''-\omega''')  \\
&=e^{\beta \omega} \Tr[e^{-\beta H}  a^\dagger a a^\dagger a]/Z - \Tr[e^{-\beta H} a^\dagger a]/Z \delta(\omega-\omega') - \Tr[e^{-\beta H} a a^\dagger]/Z \delta(\omega''-\omega''')  \\
&=\frac{1}{1-e^{\beta \omega} } \left( - \langle a^\dagger a \rangle \delta (\omega-\omega') - \langle aa^\dagger  \rangle \delta (\omega''-\omega''')\right) \\
&= \frac{e^{\beta\omega}+1} {(e^{\beta\omega}-1)^2} \delta(\omega-\omega')\delta(\omega''-\omega''') ,
\end{split}
\end{align}
where we used that $a^\dagger a$ commutes with the Hamiltonian in the first line. We have not been precise which Fourier mode we exactly track by using $\omega = \omega''$. The error is suppressed when we consider the smearing $a=\int d\omega' a_{\omega'}$, with integration bounds $(\omega-\frac{1} {2} \omega_s, \omega+\frac{1} {2} \omega_s)$. We then obtain
\begin{equation}
\langle a^\dagger \widetilde{a}^\dagger \widetilde{a} a \rangle = \frac{e^{\beta\omega}+1} {(e^{\beta\omega}-1)^2} \omega_s^2 ,
\end{equation}
and similarly
\begin{equation}
\langle  a \widetilde{a} \widetilde{a}^\dagger a^\dagger   \rangle = e^{\beta\omega}\frac{e^{\beta\omega}+1} {(e^{\beta\omega}-1)^2} \omega_s^2 .
\end{equation}
We can, therefore, compute the overlap between $\ket{\Psi_0}$ and $\ket{\Psi_1} = \widetilde{a}a\ket{\Psi_0}/{\cal N}$, where ${\cal N}$ is the normalization,
\begin{equation}
\langle \Psi_0 | \Psi_1 \rangle = \langle \widetilde{a}a \rangle /\sqrt{\frac{e^{\beta\omega}+1} {(e^{\beta\omega}-1)^2} \omega_s^2} = \frac{e^{\beta \omega/2}} {\sqrt{e^{\beta\omega}+1}} ,
\end{equation}
and similarly for the state $\ket{\Psi_1'} \sim \widetilde{a}^\dagger a^\dagger\ket{\Psi_0}$ we obtain the overlap
\begin{equation}
\langle \Psi_0 | \Psi_1' \rangle  =  \frac{1} {\sqrt{e^{\beta\omega}+1}} .
\end{equation}

We can also calculate the overlap between two states perturbed different frequencies, i.e. $\ket{\Psi_1} \sim \widetilde{a} a \ket{\Psi_0}$ and $\ket{\Psi_2} \sim \widetilde{b} b \ket{\Psi_0}$. The relevant four point function is
\begin{align}
\begin{split}
\langle b^\dagger \widetilde{b}^\dagger \widetilde{a} a \rangle &= \langle b^\dagger  a e^{-\beta H/2} a^\dagger b e^{\beta H/2}\rangle \\
&=e^{-\beta(\omega_a-\omega_b)/2} \langle b^\dagger  a  a^\dagger b \rangle \\
&=e^{-\beta(\omega_a-\omega_b)/2} \left( \Tr[e^{-\beta H} b^\dagger b a  a^\dagger]/Z\right) \\
&=e^{-\beta(\omega_a-\omega_b)/2} \left( \Tr[e^{-\beta H} b^\dagger b a^\dagger  a]/Z + \Tr[e^{-\beta H} b^\dagger b]/Z\delta(\omega_a-\omega_a') \right) \\
&=e^{-\beta(\omega_a-\omega_b)/2} \left( e^{-\beta \omega_a} \Tr[e^{-\beta H} b^\dagger b a a^\dagger  ]/Z + \langle b^\dagger b\rangle\delta(\omega_a-\omega_a') \right) \\
&=\frac{ e^{-\beta(\omega_a-\omega_b)/2} } {(1-e^{-\beta \omega_a}) (e^{\beta \omega_b} -1)}\delta(\omega_a-\omega_a')\delta(\omega_b-\omega_b') \\
&=\frac{ e^{\beta(\omega_a+\omega_b)/2} } {(e^{\beta \omega_a}-1) (e^{\beta \omega_b} -1)}\delta(\omega_a-\omega_a')\delta(\omega_b-\omega_b') , \\
\end{split}
\end{align}
which we use to calculate the overlap, after taking the smearing into account
\begin{align}
\begin{split}
\langle \Psi_2 | \Psi_1 \rangle &= \frac{ e^{\beta(\omega_a+\omega_b)/2} } {(e^{\beta \omega_a}-1) (e^{\beta \omega_b} -1)} \sqrt{\frac{(e^{\beta\omega_a}-1)^2} {e^{\beta\omega_a}+1}}  \sqrt{\frac{(e^{\beta\omega_b}-1)^2} {e^{\beta\omega_b}+1}} \\
&= \frac{ e^{\beta(\omega_a+\omega_b)/2} } {\sqrt{e^{\beta \omega_a}+1} \sqrt{e^{\beta \omega_b} +1}} .
\end{split}
\end{align}
In the limit that the frequencies are almost the same, but still in different bins, we obtain
\begin{equation} 
\langle \Psi_2 | \Psi_1 \rangle = \frac{ e^{\beta \omega}} {e^{\beta \omega}+1} .
\end{equation}
We can obtain the expression for the states $\ket{\Psi_1'} \sim \widetilde{a}^\dagger a^\dagger \ket{\Psi_0}$ and $\ket{\Psi_2'} \sim \widetilde{b}^\dagger b^\dagger \ket{\Psi_0}$ in the same limit in a similar manner
\begin{equation} 
\langle \Psi_2' | \Psi_1' \rangle = \frac{1} {e^{\beta \omega}+1} .
\end{equation}

To conclude this appendix, we note that the most important result is that the overlap is significantly not equal to one for a large range of frequencies. This is necessary for the assumption that the states are independent.

\section{Overlap of States}
\label{app:overlap}
\begin{figure}[t]
\centering
    \includegraphics[width=.8\textwidth]{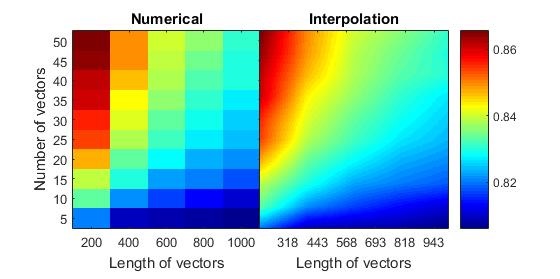}
\caption{Numerical search for the largest overlap following the method described in equation \ref{eqn:max}.}
\label{fig:overlap}
\end{figure}

In this appendix, we try to estimate how many states we need to get close to a given test state, recall equation \eqref{eqn:natconlap},
\begin{equation}
\ket{\Psi_1(t_0)} + \ket{\delta} = a_0 \ket{\Psi_0} + a_1 A_1 \widetilde{A}_1 \ket{\Psi_0} + a_2 A_2 \widetilde{A}_2 \ket{\Psi_0}...
\end{equation}
This question cannot be answered in general. Therefore, we study a simpler problem. Given a test state, how many random states do we need to get close to the test state.
\begin{equation}
\ket{\Psi(t_0)} + \ket{\delta} = a_1 \ket{\Psi_1} + a_2 \ket{\Psi_2} + a_3 \ket{\Psi_3}...
\end{equation}
We can rewrite this as a least square problem, where the matrix $A$ has the states $\ket{\Psi_i}$ as its columns and the vector $x$ has the coefficients $a_i$ as its entries.
\begin{equation}
Ax=\ket{\Psi(t_0)} ,
\end{equation}
and maximize over the phases after obtaining the least squares solution. Maximizing over the phases is a computational hard problem and we, therefore, rephrase the problem as follows. We instead maximize
\begin{equation}
M=\max_{t_0} ( \bra{\Psi(t_0)}Ax),
\end{equation}
first over the phases, which we can do by taking the elementwise absolute value, 
\begin{equation}
\label{eqn:max}
M=|\bra{\Psi}|\cdot |Ax|,
\end{equation}
and then numerically maximize over $x$. The results are shown in figure \ref{fig:overlap}. We use interpolation to get a clearer view of how the overlap develops.

These results suggest that there is linear relation between the length of the vectors and the number of vectors in the superposition for constant overlap. However, numerical algorithms can only find local extremes, and by repeating this many times an estimate for the global minimum is obtained. The results are, therefore, too small. This effect becomes stronger with a larger parameter space (more vectors in the superposition) and a more rapidly changing function (longer vectors).

\section{Volume of Self-Dual Cones}
\label{app:cones}
In this appendix, we will discuss some shapes of self dual cones and how they affect the discussion about the natural cones. We will restrict ourselves to Hilbert spaces of the form $\mathbb{R}^n$ to simplify notation. We can always restrict to the real subspace that a cone lies on in a complex Hilbert space. We repeat the definition of the dual of a cone as a reminder
\begin{equation}
K^D \coloneqq \{ \ket{x} \in {\cal H} : \braket{x | y} \geq 0, \forall \ket{y} \in K  \} .
\end{equation}
The volume of the natural cone is difficult to estimate. However, the volume of the natural cone is constrained to between the orthant cone and the Lorentz cone, which are the two examples that we will discuss.

\subsection{The Orthant Cone}
The most straightforward cone is the orthant cone. This the cone that has non-negative amplitudes in some basis.
\begin{equation}
K_O \coloneqq \{ \ket{x} \in {\cal H}\quad |\quad\ket{x}=\sum_i a_i \ket{i}, a_i \geq 0\} ,
\end{equation} 
where $\ket{i}$ are the basis vectors. It is trivial to show that this is a cone, that it is convex, and that it is self-dual. We can generate a new cone by flipping the sign of one (or more) of the basis vectors, and cover the Hilbertspace ${\cal H}$ by doing this in all possible combinations, i.e. by $2^n$ orthant cones. This means that the volume of one of these cones is given by
\begin{equation}
\frac{\text{Vol} (K_O \cap S^{n-1})} {\text{Vol} ({\cal H} \cap S^{n-1} )} = 2^{-n} ,
\end{equation}
where $n$ is the dimension of the Hilbert space and $S^{n-1}$ is the $(n-1)$-sphere to restrict to normalized states.

\subsection{The Circular Cone}
Another example of a commonly discussed cone is the circular cone.
\begin{equation}
K_{\ket{u},\theta} \coloneqq \{ \ket{x} \in {\cal H}\quad |\quad 	\measuredangle (\ket{x},\ket{u}) \leq \theta\} ,
\end{equation}
for some axis of rotation $\ket{u}$ and angle $\theta$. The dual of this cone is given by
\begin{equation}
K_{\ket{u},\theta}^D = K_{\ket{u},\frac{\pi} {2} - \theta} .
\end{equation}
The cone is, therefore, self-dual for the angle $\theta=\pi/4$. This coincides with the orthant cone in two dimensions.

The self-dual circular cone is called the Lorentz cone.
\begin{equation}
K_L\coloneqq \{ (\ket{x},t) \in {\cal H}\quad |\quad ||x||_2 \leq t\} ,
\end{equation}
where we could use any basis vector for $t$, this direction is excluded from $\ket{x}$.

The volume of a circular cone is the area of a spherical cap, which for the case of $\theta=\pi/4$, is given by
\begin{equation}
\frac{\text{Vol} (K_{\ket{u},\pi/4} \cap S^{n-1} )} {\text{Vol} ({\cal H} \cap S^{n-1} )} = \frac{1} {2} I_{1/2} \left(\frac{n-1} {2}, \frac{1} {2}\right) \propto n^{-1/2} (\sqrt{2})^{-n} ,
\end{equation}
where $I_x(a,b)$ is the regularized incomplete beta function.

\bibliographystyle{jhep}
\bibliography{references}

\providecommand{\href}[2]{#2}\begingroup\raggedright\begin{thebibliography}{10}

\bibitem{Hawking:1974sw}
S.~Hawking, \emph{{Particle Creation by Black Holes}},
  \href{https://doi.org/10.1007/BF02345020}{\emph{Commun.Math.Phys.} {\bfseries
  43} (1975) 199}.

\bibitem{Mathur:2009hf}
S.~D. Mathur, \emph{{The Information paradox: A Pedagogical introduction}},
  \href{https://doi.org/10.1088/0264-9381/26/22/224001}{\emph{Class.Quant.Grav.}
  {\bfseries 26} (2009) 224001}
  [\href{https://arxiv.org/abs/0909.1038}{{\ttfamily 0909.1038}}].

\bibitem{Almheiri:2013hfa}
A.~Almheiri, D.~Marolf, J.~Polchinski, D.~Stanford and J.~Sully, \emph{{An
  Apologia for Firewalls}},
  \href{https://doi.org/10.1007/JHEP09(2013)018}{\emph{JHEP} {\bfseries 1309}
  (2013) 018} [\href{https://arxiv.org/abs/1304.6483}{{\ttfamily 1304.6483}}].

\bibitem{Marolf:2013dba}
D.~Marolf and J.~Polchinski, \emph{{Gauge/Gravity Duality and the Black Hole
  Interior}},
  \href{https://doi.org/10.1103/PhysRevLett.111.171301}{\emph{Phys.Rev.Lett.}
  {\bfseries 111} (2013) 171301}
  [\href{https://arxiv.org/abs/1307.4706}{{\ttfamily 1307.4706}}].

\bibitem{Maldacena:1997re}
J.~M. Maldacena, \emph{{The Large N limit of superconformal field theories and
  supergravity}},
  \href{https://doi.org/10.1023/A:1026654312961}{\emph{Adv.Theor.Math.Phys.}
  {\bfseries 2} (1998) 231}
  [\href{https://arxiv.org/abs/hep-th/9711200}{{\ttfamily hep-th/9711200}}].

\bibitem{Papadodimas:2012aq}
K.~Papadodimas and S.~Raju, \emph{{An Infalling Observer in AdS/CFT}},
  \href{https://doi.org/10.1007/JHEP10(2013)212}{\emph{JHEP} {\bfseries 1310}
  (2013) 212} [\href{https://arxiv.org/abs/1211.6767}{{\ttfamily 1211.6767}}].

\bibitem{Papadodimas:2013b}
K.~Papadodimas and S.~Raju, \emph{{The Black Hole Interior in AdS/CFT and the
  Information Paradox}},  \href{https://arxiv.org/abs/1310.6334}{{\ttfamily
  1310.6334}}.

\bibitem{Papadodimas:2013}
K.~Papadodimas and S.~Raju, \emph{{State-Dependent Bulk-Boundary Maps and Black
  Hole Complementarity}},  \href{https://arxiv.org/abs/1310.6335}{{\ttfamily
  1310.6335}}.

\bibitem{deBoer:2018ibj}
J.~de~Boer, R.~Van~Breukelen, S.~F. Lokhande, K.~Papadodimas and E.~Verlinde,
  \emph{{On the interior geometry of a typical black hole microstate}},
  \href{https://doi.org/10.1007/JHEP05(2019)010}{\emph{JHEP} {\bfseries 05}
  (2019) 010} [\href{https://arxiv.org/abs/1804.10580}{{\ttfamily
  1804.10580}}].

\bibitem{deBoer:2019kyr}
J.~De~Boer, R.~Van~Breukelen, S.~F. Lokhande, K.~Papadodimas and E.~Verlinde,
  \emph{{Probing typical black hole microstates}},
  \href{https://arxiv.org/abs/1901.08527}{{\ttfamily 1901.08527}}.

\bibitem{Papadodimas:2015xma}
K.~Papadodimas and S.~Raju, \emph{{Local Operators in the Eternal Black Hole}},
  \href{https://doi.org/10.1103/PhysRevLett.115.211601}{\emph{Phys. Rev. Lett.}
  {\bfseries 115} (2015) 211601}
  [\href{https://arxiv.org/abs/1502.06692}{{\ttfamily 1502.06692}}].

\bibitem{vanBreukelen:2017dul}
R.~van Breukelen and K.~Papadodimas, \emph{{Quantum teleportation through
  time-shifted AdS wormholes}},
  \href{https://doi.org/10.1007/JHEP08(2018)142}{\emph{JHEP} {\bfseries 08}
  (2018) 142} [\href{https://arxiv.org/abs/1708.09370}{{\ttfamily
  1708.09370}}].

\bibitem{Maldacena:2001kr}
J.~M. Maldacena, \emph{{Eternal black holes in anti-de Sitter}}, {\emph{JHEP}
  {\bfseries 0304} (2003) 021}
  [\href{https://arxiv.org/abs/hep-th/0106112}{{\ttfamily hep-th/0106112}}].

\bibitem{Gao:2016bin}
P.~Gao, D.~L. Jafferis and A.~Wall, \emph{{Traversable Wormholes via a Double
  Trace Deformation}},  \href{https://arxiv.org/abs/1608.05687}{{\ttfamily
  1608.05687}}.

\bibitem{Maldacena:2017axo}
J.~Maldacena, D.~Stanford and Z.~Yang, \emph{{Diving into traversable
  wormholes}}, \href{https://doi.org/10.1002/prop.201700034}{\emph{Fortsch.
  Phys.} {\bfseries 65} (2017) 1700034}
  [\href{https://arxiv.org/abs/1704.05333}{{\ttfamily 1704.05333}}].

\bibitem{Bratteli:1979tw}
O.~Bratteli and D.~W. Robinson, \emph{{Operator algebras and quantum
  statistical mechanics. 1. C* and W* algebras, symmetry groups, decomposition
  of states.}} 1979.

\bibitem{Papadodimas:2017qit}
K.~Papadodimas, \emph{{A class of non-equilibrium states and the black hole
  interior}},  \href{https://arxiv.org/abs/1708.06328}{{\ttfamily 1708.06328}}.

\bibitem{Bousso:2013wia}
R.~Bousso, \emph{{Firewalls From Double Purity}},
  \href{https://arxiv.org/abs/1308.2665}{{\ttfamily 1308.2665}}.

\bibitem{lloyd}
S.~{Lloyd}, \emph{{Pure state quantum statistical mechanics and black holes}},
  {\emph{ArXiv e-prints} (2013) }
  [\href{https://arxiv.org/abs/1307.0378}{{\ttfamily 1307.0378}}].

\bibitem{Shenker:2013yza}
S.~H. Shenker and D.~Stanford, \emph{{Multiple Shocks}},
  \href{https://doi.org/10.1007/JHEP12(2014)046}{\emph{JHEP} {\bfseries 12}
  (2014) 046} [\href{https://arxiv.org/abs/1312.3296}{{\ttfamily 1312.3296}}].

\bibitem{'tHooft:1984re}
G.~'t~Hooft, \emph{{On the Quantum Structure of a Black Hole}},
  \href{https://doi.org/10.1016/0550-3213(85)90418-3}{\emph{Nucl.Phys.}
  {\bfseries B256} (1985) 727}.

\bibitem{Harlow:2014yoa}
D.~Harlow, \emph{{Aspects of the Papadodimas-Raju Proposal for the Black Hole
  Interior}}, \href{https://doi.org/10.1007/JHEP11(2014)055}{\emph{JHEP}
  {\bfseries 11} (2014) 055} [\href{https://arxiv.org/abs/1405.1995}{{\ttfamily
  1405.1995}}].

\end{thebibliography}\endgroup
\end{document}